\documentclass[Afour,sageh,times]{sagej}
\usepackage{moreverb,url}

\usepackage[colorlinks,bookmarksopen,bookmarksnumbered,citecolor=red,urlcolor=red]{hyperref}

\newcommand\BibTeX{{\rmfamily B\kern-.05em \textsc{i\kern-.025em b}\kern-.08em
		T\kern-.1667em\lower.7ex\hbox{E}\kern-.125emX}}

\def\volumeyear{2025}

\usepackage{amsthm}
\usepackage{natbib}
\setcitestyle{numbers,square}

\usepackage{amsmath} % assumes amsmath package installed
\usepackage{amssymb}  % assumes amsmath package installed
\usepackage{algorithmic}
\usepackage{algorithm}

\usepackage{graphicx}
\usepackage{textcomp}
\usepackage{xcolor}

\theoremstyle{definition}
\newtheorem{remark}{Remark}[]

\begin{document}
\runninghead{Ke}

\title{A Fast Initialization Method for Neural Network Controllers: A Case Study of Image-based Visual Servoing Control for the multicopter Interception}

\author{Chenxu Ke\affilnum{2}, Congling Tian\affilnum{2}, Kaichen Xu\affilnum{1}, Ye Li\affilnum{2} and Lingcong Bao\affilnum{2}}

\affiliation{
	\affilnum{1}State Key Laboratory of Fluid Power and Mechatronic Systems, School of Mechanical Engineering, Zhejiang University, Hangzhou, China\\
	\affilnum{2}Zhiyuan Research Institute, Hangzhou, China}

\corrauth{Chenxu Ke
	Zhiyuan Research Institute,
	Hangzhou, China.}

\email{kcx064@zju.edu.cn}
%\maketitle
%\thispagestyle{empty}
%\pagestyle{empty}

%TODO 摘要需要结合题目重写，引言需要调整为与新题目适配的状态。重点都是神经网络控制器的快速生成。基于强化学习的神经网络控制器难以实现快速训练和落地部署，且缺少稳定性保证。因此基于Lyapunov神经网络控制器被提出并取得相关研究成果，但是这类方法需要一个初始控制器网络后，再进行后续训练。然而获取初始控制器依旧需要基于控制理论进行设计，这种设计流程要求控制器设计者具有一定经验，无疑提高了应用神经网络控制器的门槛。针对获取初始稳定的神经网络控制器这一问题，本文提出了一种快速神经网络控制器初始化方法，该方法XXX。为验证该方法的可行性，本文以多旋翼飞行器为例进行说明。
%TODO 摘要、引言、结论要重写，其他部分可以结合RAL的审稿意见略加补充
\begin{abstract}
% 常见的基于强化学习的控制器设计方法在训练初期通常需要大量数据，且训练过程随机性较强、收敛速度慢，往往需要耗费大量时间或依赖较高的算力资源。
% 另一类基于学习的方法通过结合Lyapunov稳定性理论，能够获得具有稳定性保证的控制器。然而，这类方法通常要求以一个已经具备稳定性的神经网络控制器作为初始条件。
% 显然，一个稳定的神经网络控制器不仅可作为强化学习的初始策略，以专注于提升控制器性能，还可作为基于Lyapunov方法的控制器的初始状态
% 虽然基于传统控制理论可以设计出稳定的控制器，但在应对更为复杂的控制问题时，仍需设计者具备相当丰富的控制设计经验。
% 本文提出的神经网络快速初始化方法，通过基于先验模型构建符合稳定性条件的数据集，实现对神经网络控制器的初始化训练。
% 该方法不依赖任何真实的数据集。
% 同时，本文以多旋翼飞行器拦截问题为例，开展了所提方法的仿真与实验研究，验证了其有效性与实际性能。
Reinforcement learning-based controller design methods often require substantial data in the initial training phase. Moreover, the training process tends to exhibit strong randomness and slow convergence. It often requires considerable time or high computational resources.
Another class of learning-based method incorporates Lyapunov stability theory to obtain a control policy with stability guarantees. However, these methods generally require an initially stable neural network control policy at the beginning of training.
Evidently, a stable neural network controller can not only serve as an initial policy for reinforcement learning, allowing the training to focus on improving controller performance, but also act as an initial state for learning-based Lyapunov control methods.
Although stable controllers can be designed using traditional control theory, designers still need to have a great deal of control design knowledge to address increasingly complicated control problems.
The proposed neural network rapid initialization method in this paper achieves the initial training of the neural network control policy by constructing datasets that conform to the stability conditions based on the system model.
Furthermore, using the image-based visual servoing control for multicopter interception as a case study, simulations and experiments were conducted to validate the effectiveness and practical performance of the proposed method.
In the experiment, the trained control policy attains a final interception velocity of 15 m/s.
%This approach does not rely on any real-world datasets.
\end{abstract}

%\begin{keywords}
%	learning-based Lyapunov control, visual servoing, high-speed interception, D-learning
%\end{keywords}

\maketitle
%%%%%%%%%%%%%%%%%%%%%%%%%%%%%%%%%%%%%%%%%%%%%%%%%%%%%%%%%%%%%%%%%%%%%%%%%%%%%%%%
\section{INTRODUCTION}
In recent years, an increasing number of studies have utilized learning-based methods to address control issues\cite{ConnellNeuralFlyenables2022}. There are two types of learning-based methods for control problem. 
One is the reinforcement learning (RL) method, and the other is the learning-based Lyapunov control (LLC) method.
In contrast to conventional control methods, RL techniques engage with the environment via trial and error to identify optimal strategies and may complete intricate tasks without dependence on exact models of the controlled entities.
RL, akin to control systems, functions through feedback mechanisms.
While RL largely uses input to refine its decision-making processes, control systems focus on achieving predetermined targets mainly by using static controller techniques during operation. 
The training process of RL can be unstable and unsafety\cite{ShenDOPTDlearning2025}, especially in safety-critical situations like the unmanned aerial vehicle (UAV) visual servoing control.
In order to ensure that the trained policy can be applied in practice, the datasets used in the training process should encompass the Region of Interest (RoI)\cite{Dai2021,DawsonSafeControlLearned2023}, which is exceedingly challenging before obtaining a available control policy. 
% 为了确保所训练的控制策略能够应用，需要用于训练的数据完全覆盖兴趣区域\cite{Dai2021, DawsonSafeControlLearned2023}，但是这样的代价是昂贵的，特别是在得到可用控制策略前。

The Lyapunov stability method provides a definitive analytical and design framework in control theory, especially for nonlinear systems \cite{LYAPUNOV1992}.
Numerous research studies have recently integrated Lyapunov stability approaches into learning-based control, referred to as Lyapunov function learning, thereby providing formal stability guarantees for deep neural network policies.
In the studies \cite{HanActorCriticReinforcement2020,DuReinforcementLearningSafe2023}, the Lyapunov function is utilized as a critic function to assess policies performance.
In \cite{Chowlyapunovbasedapproach2018} and \cite{ChowLyapunovbasedsafe2019}, Lyapunov functions are integrated into optimization frameworks to guarantee system stability.
The Lyapunov stability condition is incorporated into the reward design in \cite{GanaiLearningStabilizationControl2023} and \cite{ChangStabilizingNeuralControl2021}.
References \cite{GanaiLearningStabilizationControl2023} and \cite{ChangStabilizingNeuralControl2021} develop the target control policy by incorporating a Lyapunov function into the reward design.
The research in \cite{QuanControlPatternsD2025} proposes learning the Lyapunov function and its derivative (referred to as the D-function) from expert demonstration data while adhering to stability constraints, thus facilitating the development of a control policy that inherently ensures Lyapunov stability.
Note that uniformly sampled data is necessary for this approach.
Otherwise, the D-function employed may not accurately represent the actual system model.
While these methods offer formal stability guarantees for the target policy and yield favorable outcomes, they depend on the \textbf{posteriori expert controllers} or trajectories and are not suitable for the original design of the control policy.

The posteriori expert controllers are also used as the initial policy of RL to circumvent the drawback of slow convergence at the beginning of the training.
Although the conventional control theory can be applied for the controller designing, the rich experience of that is also important to solve a complex control problem.
\textbf{Therefore, this paper proposed an initial policy training method that involves constructing datasets that meet stability requirements and then training a neural network control policy based on the datasets.}
Moreover, acquiring a group of datasets without a stable control policy is exceedingly challenging.
Conversely, without a controller, acquiring the model of controlled objects may be more attainable than gathering data.
Utilizing the datasets produced by mathematical models that adhere to Lyapunov stability, an untrained neural network may be directly developed into a control policy, circumventing the conventional control design and debugging procedures. 
The trained neural network control policy can be enhanced further by RL methods or the LLC method that needs an initial stable control policy \cite{ShenDOPTDlearning2025}.
In this paper, the case of image-based visual servoing control for the multicopter interception is adopted to demonstrate the effectiveness of the proposed method. 
In the experiment, a final flight speed of up to 15 m/s was achieved. 
Noted that the purpose of this method is to obtain a usable control policy with mediocre performance at least, and the optimization of the control policy still needs to be accomplished through RL and LLC methods.

The paper is organized as follows: Section II outlines the coordinate systems and mathematical models employed. Section III presents the rapid methodology for training neural network policies, applies it to the design of multicopter interception control, and validates the stability of the trained policies by the almost Lyapunov condition\cite{LiuAlmostLyapunovfunctions2020}. Section IV presents the experimental result of the trained policies implemented on quadrotor platforms.

\begin{figure}[!t]
	\centering
	\includegraphics[width=8.5cm]{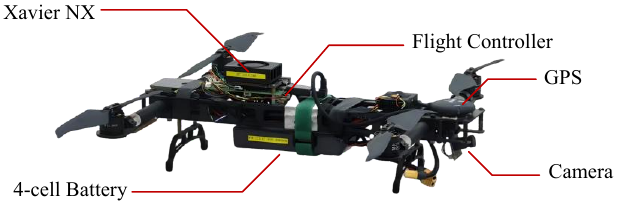}
	\caption{The quadcopter used for the target interception has the camera installed with a 15° pitch angle.}
	\label{quadcopter}
\end{figure}
\section{PROBLEM FORMULATION}
% 总分
% 本文在基于图像的视觉伺服框架下，基于D学习方式快速设计无人机拦截算法

In this paper, the proposed method will be used to obtain the image-based visual servoing policy for multicopter interception.
The policy is a neural network with the target position in the image plane coordinates as the input. 
Here, the desired position of the target point is the center of the image plane\cite{YangHighSpeedInterception2025,YangAutonomousInterceptDrone2020}.
\begin{figure}[!t]
	\centering
	%  \framebox{\parbox{3.3in}{\includegraphics[width=8.3cm]{img/ContourPlot.pdf}}}
	\includegraphics[width=8.5cm]{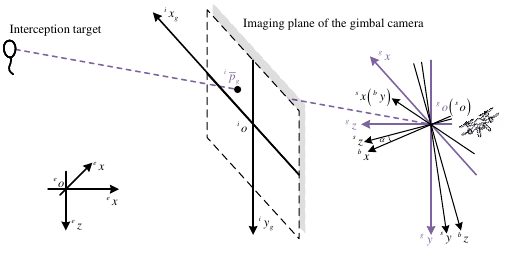}
	\caption{Description of coordinate systems in the multicopter interception problem. The left superscript indicates the abbreviation of the coordinate system, where the letter `g' represents the gimbal camera coordinate system, the letter `e' represents the Earth-fixed coordinate system, the letter `s' represents the strapdown camera coordinate system, and the letter `b' represents the body coordinate system. The installation angle of the strapdown camera is $\alpha$. The dashed rectangle represents the imaging plane of the camera. ${^i}p$ is the position of the interception target in the image coordinate system.}
	\label{CoordinateFrame}
\end{figure}

%In this section, image-based visual servoing multicopter interception model is proposed. With them, the multicopter interception control design problem is formulated.

\subsection{Coordinate systems and models}

This paper uses five coordinate systems, which are the Earth-fixed coordinate system (EFCS), the body coordinate system (BCS), the gimbal camera coordinate system (GCCS), the strapdown camera coordinate system (SCCS), and the image coordinate system (ICS), as shown in Fig.\ref{CoordinateFrame}. The positive orientation adheres to the right-hand principle.
The Earth-fixed coordinate axes $\left\{{^e}x,{^e}y,{^e}z\right\}$ point north, east, and downward to gravity, respectively. The body coordinate axes $\left\{{^b}x,{^b}y,{^b}z\right\}$ point forward, rightward, and downward of the multicopter, respectively. 
Assuming the camera is connected to the multicopter via a two-axis gimbal (including roll axis and pitch axis), the $\left\{{^g}x,{^g}y,{^g}z\right\}$ axes of GCCS point to the rightward, downward, and forward directions of the camera's optical axis, respectively.
The GCCS and BCS share the same origin. The orientations of BCS and GCCS frames differ solely by a tilt angle, which can be decomposed into pitch and roll angles for clarity. In this study, the camera and aircraft are mounted directly without a gimbal, which makes the system more reliable \cite{YangHighSpeedInterception2025}. This paper describes the installation attitude of the strapdown camera with reference to BCS. The strapdown camera is installed with only a non-zero pitch angle ($\alpha\neq0$) as shown in Fig.\ref{CoordinateFrame}, and the positive direction is likewise defined by the right-hand rule. To improve the interception capability of the strapdown camera aircraft, it is necessary to set $\alpha>0$\cite{YangLineSightConstrained2025}. 
Compared with strapdown cameras, gimbal cameras decouple the line of sight from the aircraft's tilt angle, thereby simplifying the interception problem addressed in this paper. 
Accordingly, based on principles of camera imaging geometry, an equivalent transformation model is established to convert image coordinates from the strapdown camera to the gimbal camera.Similar to the GCCS, the defined SCCS coordinate axes $\left\{{^s}x,{^s}y,{^s}z\right\}$ point rightward, downward, and forward along the camera's optical axis, respectively. 
The transformation between these two coordinate systems can be fully described by the roll angle $\phi_s$ and pitch angle $\theta_s$ of the strapdown camera relative to EFCS. The image coordinate axes $\left\{{^i}x,{^i}y\right\}$ point rightward and downward, respectively. The origin of ICS is the center of the image. The subscript $s$ denotes the ICS axes and coordinates of the strapdown camera, while the subscript $g$ represents the ICS axes and coordinates of the gimbal camera.

\begin{figure}[!t]
	\centering
	\includegraphics[width=8.0cm]{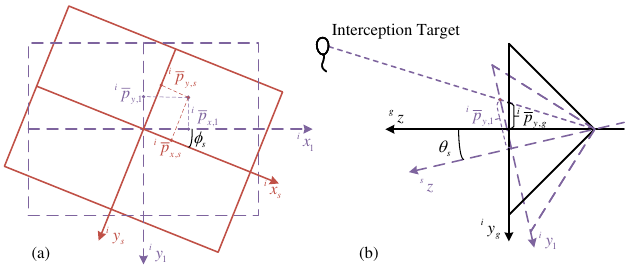}
	\caption{Geometric relationship between the target point's coordinates in the strapdown camera image and its coordinates in the gimbal camera image}
	\label{imagecoordinate}
\end{figure}
Assuming that the coordinate of the interception target in ICS of the strapdown camera is denoted as $\left({^i}{\overline{p}}_{x,s},{^i}{\overline{p}}_{y,s}\right)$, which can be converted to ICS of the gimbal camera by roll angle $\phi_s$ and pitch angle $\theta_s$ of the strapdown camera. 
As shown in Fig.\ref{imagecoordinate}(a), $\left({^i}{\overline{p}}_{x,1},{^i}{\overline{p}}_{y,1}\right)$ is a temporary coordinate value computed based on the roll angle $\phi_s$. 
That is
\begin{equation}\label{convert1}
	\begin{aligned}
		{^i}\overline{p}_{x,1} &= {^i}\overline{p}_{x,s}\cos\phi_s - {^i}\overline{p}_{y,s}\sin\phi_s\\
		{^i}\overline{p}_{y,1} &= {^i}\overline{p}_{x,s}\sin\phi_s + {^i}\overline{p}_{y,s}\cos\phi_s\\
	\end{aligned} 
\end{equation}
According to Fig.\ref{imagecoordinate}(b), then the coordinate in ICS of the gimbal camera is derived as
\begin{equation}\label{convert2}
	\begin{aligned}
		{^i}\overline{p}_{x,g} &= {^i}\overline{p}_{x,1}\\
		{^i}\overline{p}_{y,g} &= \tan\left(\arctan {^i}\overline{p}_{y,1} - \theta_s\right)
	\end{aligned}
\end{equation}
Then, $\left({^i}\overline{p}_{x,g},{^i}\overline{p}_{y,g}\right)$ is the coordinate of the interception target in ICS of the gimbal camera.

\begin{remark}
	The coordinate symbol with overline is the normalized values. Take the coordinate symbol of ICS of strapdown camera, namely $\left({^i}{\overline{p}}_{x,{s}},{^i}{\overline{p}}_{y,{s}}\right)$, as example, it is calculated by 
	\begin{equation}
		{^i}{\overline{\mathbf{p}}}_{s} = \left[{^i}\overline{p}_{x,{s}},{^i}\overline{p}_{y,\mathrm{s}}\right]^\mathrm{T} = \left[\frac{{^i}p_{x,{s}}}{f_\mathrm{oc}}, \frac{{^i}p_{y,{s}}}{f_\mathrm{oc}}\right]^\mathrm{T}
	\end{equation}
	where $f_\mathrm{oc}$ is focal of length in pixel units, and $\left({^i}p_{x,\mathrm{s}},{^i}p_{y,{s}}\right)$ is the coordinates of the image point expressed in pixel units. Similarly, the coordinates of the target point in the image plane of the gimbal camera are simply expressed as
	\begin{equation}
		{^i}\overline{\mathbf{p}}_g \triangleq \left({^i}\overline{p}_{x}, {^i}\overline{p}_{y}\right)
	\end{equation} 
\end{remark}

% TODO: 备用内容+Remark转换到云台坐标系需要捷联相机的横滚角和俯仰角，这两的角度需要由机体横滚俯仰角加上捷联相机的安装角计算得到。目标在捷联相机图像中的坐标位置可以等效转换到云台相机图像中的坐标.

% 本文的目标是得到一个控制器使得，无人机在指定前飞行速度下能够基于图像数据逼近被拦截目标，从而实现拦截。换句话说，需要趋势无人机使得目标位于图像中的指定位置。直观的来讲，目标在图像中的位置，取决于摄像机自身的速度和角速度。因此，这里将给出目标在图像中位置与相机自身速度和角速度的关系。
%The objective of this paper is to design a controller that enables the unmanned aerial vehicle (UAV) to intercept a target by approaching it based on image data while maintaining a specified forward flight speed. Specifically, the controller will guide the UAV such that the target is positioned at a designated location within the camera's field of view. Intuitively, the position of the target in the image is influenced by both the translational and angular velocities of the camera. Therefore, this paper will derive the relationship between the target's position in the image and the camera's velocity and angular velocity.(tran)

The dynamics of the target point $^i\overline{\mathbf{p}}_{g}$ in the gimbal camera image plane\cite{Chaumette_2006} is
\begin{equation}\label{imageJacobianMatrixEq}
	{^i}\dot{\overline{\mathbf{p}}}_{g} = \mathbf{L}_\mathbf{s} {^\mathrm{g}}\tilde{\mathbf{v}}
\end{equation}
where the image Jacobian matrix $\mathbf{L}_\mathbf{s}$ is  
\begin{equation}\label{imageJacobianMatrix}
	\mathbf{L}_\mathbf{s} = \begin{bmatrix}
		-\frac{1}{c_z} & 0 & \frac{{^i}\overline{p}_{x}}{c_z}& {^i}\overline{p}_{x}{^i}\overline{p}_{y} & -(1+{^i}\overline{p}_x^2) &{^i}\overline{p}_{y}\\
		0 & -\frac{1}{c_z} & \frac{{^i}\overline{p}_y}{c_z}& 1+{^i}\overline{p}_y^2 & -{^i}\overline{p}_x{^i}\overline{p}_y      &-{^i}\overline{p}_x\\
	\end{bmatrix},
\end{equation}
$c_z$ is the object distance of the gimbal camera,
\begin{equation}
	{^\mathrm{g}}\tilde{\mathbf{v}} = \left[{^\mathrm{g}}\mathbf{v}^\mathrm{T}\ {^\mathrm{g}}\boldsymbol{\omega}^\mathrm{T}\right]^\mathrm{T},
\end{equation}
${^\mathrm{g}}\mathbf{v} = \left[{^\mathrm{g}}v_x\ {^\mathrm{g}}v_y\ {^\mathrm{g}}v_z\right]^\mathrm{T}$ is the velocity of the gimbal camera in GCCS, and
${^\mathrm{g}}\boldsymbol{\omega} = \left[{^\mathrm{g}}\omega_x\ {^\mathrm{g}}\omega_y\ {^\mathrm{g}}\omega_z\right]^\mathrm{T}$ is the angular rate of the gimbal camera in GCCS. In this paper, the coordinates of the target point in the image plane of the gimbal camera are unaffected by the pitch and roll motion of the body, that is ${^g}\omega_x = 0$ and ${^g}\omega_z = 0$.
Then Eq.\eqref{imageJacobianMatrixEq} is simplified as 
\begin{equation}
	\label{small_interception_model}
	{^i}\dot{\overline{\mathbf{p}}} = \mathbf{L}_\mathbf{s}^{'} {^\mathrm{g}}\tilde{\mathbf{v}}^{'}
\end{equation}
where the image Jacobian matrix $\mathbf{L}_\mathbf{s}^{'}$ is
\begin{equation}\label{imageJacobianMatrix_small}
	\mathbf{L}_\mathbf{s}^{'} = \begin{bmatrix}
		-\frac{1}{c_z} & 0 & \frac{{^i}\overline{p}_x}{c_z} &-(1+{^i}\overline{p}_x^2)\\
		0 & -\frac{1}{c_z} & \frac{{^i}\overline{p}_y}{c_z} &-{^i}\overline{p}_x{^i}\overline{p}_y\\
	\end{bmatrix},
\end{equation}
and
\begin{equation}
	{^\mathrm{g}}\tilde{\mathbf{v}}^{'} = \left[{^\mathrm{g}}v_x\ {^\mathrm{g}}v_y\ {^\mathrm{g}}v_z\ {^\mathrm{g}}\omega_y\right]^\mathrm{T}.
\end{equation}

%TODO For simplicity, multicopter is modeled as a rigid body accepted velocity command ${^\mathrm{e}}\mathbf{v}_\mathrm{d}$, that is 
%\begin{equation}
%	\begin{aligned}
%		{^\mathrm{e}}\dot{\mathbf{p}} &= {^\mathrm{e}}\mathbf{v}\\
%		\epsilon{^\mathrm{e}}\dot{\mathbf{v}} &= -{^\mathrm{e}}\mathbf{v} + {^\mathrm{e}}\mathbf{v}_\mathrm{d},
%	\end{aligned}
%\end{equation}
%where $\epsilon$ is the time constant of body velocity, ${^\mathrm{e}}\mathbf{p}$ and ${^\mathrm{e}}\mathbf{v}$ are the position and velocity of the multicopter in EFCS, respectively.

\subsection{Control strategy}\label{control_policy}
%为了实现多旋翼飞行器拦截任务，本研究采用的策略是：令飞行器以指定速度沿着云台相机${^g}z$轴方向飞行，同时由神经网络控制器控制飞行器沿云台相机的${^g}x$轴和${^g}y$轴方向的运动速度，使得被拦截对象在云台相机图像中位置趋于图像中心。

In order to accomplish the interception mission with the multicopter, the strategy employed in this research is as follows:
The multicopter is made to fly at a designated velocity $v_{z,g}$ along the ${^g}z$-axis, which is a breeze for multicopters.
Simultaneously, the designed policy, namely controller, is utilized to regulate the motion velocities of the multicopter along the ${^g}x$-axis and ${^g}y$-axis so that the position of the intercepted object in the gimbal camera image tends to the center of the image. 
Note that the position of the intercepted object in the gimbal camera image is converted from that in the strapdown camera image through Eq.\eqref{convert1} and Eq.\eqref{convert2}.

%\begin{itemize}
%
%\item 
%\item 
%\item 
%\item 
%
%\end{itemize}

\section{POLICY GENERATE}
%在本节，将给出基于神经网络的拦截控制器设计方法。
%首先简单介绍控制器设计过程，然后就多旋翼飞行器拦截目标这一具体问题，介绍控制设计过程，最后采用几乎李雅普诺夫稳定性的对所训练的控制器进行稳定性验证。

In this section, the method to generate a neural network policy for interception will be introduced. Firstly, the policy design process will be briefly described. Secondly, a neural network policy used for object interception is designed and trained in detail. Finally, the almost Lyapunov condition\cite{LiuAlmostLyapunovfunctions2020} is adopted for control stability verification. 

\subsection{Policy training method}
%\begin{algorithm}[!h]
%	\caption{Initial controller generation algorithm with D-learning}
%	\label{alg1}
%	\begin{algorithmic}[1]  
%		\REQUIRE An untrained neural network ${c}_{\theta_0}$, system dynamic model $\dot{x}=f(x,u)$
%		\ENSURE $c_{\theta_k}(x)$
%		\STATE Design candidate Lyapunov function $V(x)$
%		\STATE Calculate the derivative of $V(x)$, that is $\frac{\partial V}{\partial x}f(x,u)$
%		\STATE Let $D(x,u) = \frac{\partial V}{\partial x}f(x,u)$ 
%		\STATE Based on the model $\dot{x}=f(x,u)$, search date point $\left(\dot{x},x\right)$ satisfying $D(x,u)\leqslant -\eta W(x)$ to construct the dataset $\mathcal{P}$, where $\eta$ is a positive number
%		\STATE Train ${c}_{\theta}$ by minimizing $L_c$ as shown in Equation \eqref{castfcn_controller_in_practice}
%	\end{algorithmic}
%\end{algorithm}
\subsubsection{Initialization}
Controller designers should initialize an untrained neural network policy $c_{\theta_0}$ according to specific control problem, where $\theta$ is the weight of neural network $c_{\theta_0}$. Meanwhile, the control object model should be known. Designers are allowed to used the object model for classic controller design. Here, the model is expressed as 
\begin{equation}
	\label{abstractmodel}
	\dot{\mathbf{x}} = f(\mathbf{x},\mathbf{u}),
\end{equation}
where $\mathbf{x}$ is the system state and the $\mathbf{u}$ is system input.
To minimize the modeling cost, the artificially constructed models often overlook the physical factors that exert a relatively minor influence on the controlled system. Consequently, the established models do not precisely mirror the actual scenario. Nevertheless, developing models grounded in objective physical laws retains significant practical relevance for control design.
%出于降低建模代价的目的，人工建立的模型会忽略掉对被控系统影响较小的物理因素，这就导致建立的模型无法与实际情况完全相同，但是基于客观物理规律建立模型对于控制设计依旧具有很强的利用价值。
%TODO：这部分内容引言中可能已有：对比表格。基于模型进行强化学习，实时交互对控制器进行训练，训练速度可能会很长和训练结果缺少稳定性依据，这类方法称为On-Policy方法。相对地，基于离线数据的Off-policy训练方法，则通过一些手段先获取训练数据集，然后开展控制器训练，这种方法需要依赖初始可用的控制器才能开展。本文方法同样基于模型，但是首先基于模型离线获取数据，然后开展控制器训练，充分利用人类对被控对象的先验知识，加速控制器的初步设计。
\subsubsection{Candidate D-function calculation}
Based on the control object, design a Lyapunov function $V(\mathbf{x})$. Then, the controlled system model \eqref{abstractmodel} is used to obtain the derivative of the $V(\mathbf{x})$, namely D-function\cite{QuanControlPatternsD2025} 
\begin{equation}\label{abstratdfunction}
	D(\mathbf{x},\mathbf{u}) = \frac{\partial V}{\partial \mathbf{x}}f(\mathbf{x},\mathbf{u}).
\end{equation}

%要具体不要泛泛而谈
\subsubsection{Searching datasets satisfies Lyapunov conditions}
For all the given $\mathbf{x}_j$ in the RoI, find corresponding $\mathbf{u}_j$ satisfying 
\begin{equation}\label{searchingdatacondition}
	\begin{aligned}
		\min_{\eta>0,\mathbf{u}_j} & \left| D(\mathbf{x}_j,\mathbf{u}_j) + \eta W(\mathbf{x}_j) \right| \\
		\mathrm{s.t.} & D(\mathbf{x}_j, \mathbf{u}_j)<0
	\end{aligned}
\end{equation} 
where $\eta$ is a positive number, $W(0) = 0$, and $W(\mathbf{x}_j)>0$ for $\forall \mathbf{x}_j\in \mathbb{R}^n \backslash \{0\}$.
The datasets 
\begin{equation}
	\mathcal{S} = \left\{\mathbf{s}_1,\mathbf{s}_2,\cdots,\mathbf{s}_N\right\}
\end{equation}
are composed of all the number of data point $\mathbf{s}_j=\left(\mathbf{x}_j,\mathbf{u}_j\right)$.

\subsubsection{Training neural network $c_{\theta_0}$ with $\mathcal{S}$}
Train ${c}_{\theta_0}$ by minimizing $L(\theta_k)$ as shown in Eq.\eqref{abstractcastfcn}
\begin{equation}\label{abstractcastfcn}
	L(\theta_k) = \min_{c_{{\theta}_k}} \frac{1}{N}\sum_{j=1}^{N}\|c_{{\theta}_k}\left(\mathbf{x}_j\right)-\mathbf{u}_j\|_2,
\end{equation}
where $k$ is the training epoch.

%\begin{equation}\label{castfcn_controller}
%	\begin{aligned}
%		L_c &= \lambda_1 \max(D(x_j,c_{\theta_k}(x_j)),0)\\
%		&+ \frac{\lambda_2}{N}\sum_{j=1}^{N}(D(x_j,c_{\theta_k}(x_j))) + \lambda_3\|\theta_k - \theta_{k-1}\|^2\\
%	\end{aligned}
%\end{equation}

\subsection{Training of interception policy}\label{training_of_controller}
\subsubsection{Initialization}
Model \eqref{imageJacobianMatrix_small} is selected for the neural network policy design. The system state vector is 
\begin{equation}
	\left[{^i}\overline{p}_x\ {^i}\overline{p}_y\ {^g}v_z\ c_z\ {^g}\omega_y\right]^\mathrm{T}\in \mathbb{R}^5.
\end{equation}
However only ${^i}\overline{p}_x$ and ${^i}\overline{p}_y$ are concerned state for the interception task.
The system input vector $\mathbf{u}$ is 
\begin{equation}
	\mathbf{u} = \left[{^g}v_x\ {^g}v_y\right]^\mathrm{T}. 
\end{equation}
This study employs a standard feed-forward neural network architecture. The network configuration consists of an input layer with three neurons and an output layer comprising a single neuron. The architecture incorporates three hidden layers, each containing 16 neurons. The hyperbolic tangent (\texttt{tanh}) activation function is implemented between all adjacent network layers, which makes the neural network smooth\cite{NeuralLyapunovControl2019}. According to subsection \textit{Control strategy}, two separate neural networks with the same architecture are used for generating ${^g}v_x$ and ${^g}v_y$, namely $c_{\theta_{x,k}}$ and $c_{\theta_{y,k}}$.
%Edition2: A standard feedforward neural network architecture is adopted in this study. The network consists of an input layer with three nodes, three hidden layers (each containing 16 neurons), and an output layer with a single node. The hyperbolic tangent (tanh) activation function is employed between all adjacent layers of the network.
%这里选用和最基础的前馈神经网络结构，输入层对应三个输入，输出层对应一个输出。除输入层和输出层外，神经网络包含三层隐藏层，每个隐藏层均为16个节点。相邻两层神经网络之间均使用双曲正切函数作为激活函数。
\subsubsection{Candidate D-function calculation}
The key to completing the interception task is to make the target position in the gimbal camera image tend towards the image center, that is ${^i}\overline{p}_x\rightarrow0$ and ${^i}\overline{p}_y\rightarrow0$. Therefore, the Lyapunov functions for ${^i}\overline{p}_x$ and ${^i}\overline{p}_y$ are respectively designed as follows:
\begin{equation}
		\begin{aligned}
			V_x &= \frac{1}{2}{^i}\overline{p}_x\cdot {^i}\overline{p}_x,\\ 
			V_y &= \frac{1}{2}{^i}\overline{p}_y\cdot {^i}\overline{p}_y.
		\end{aligned}
\end{equation}
The finally Lyapunov function is
\begin{equation}\label{Lyapunov_V}
	V = V_x + V_y.
\end{equation}
If $V_x\rightarrow0$ and $V_y\rightarrow0$, then it has $V\rightarrow0$. 
Here, the policies along ${^g}x$-axis and ${^g}y$-axis are designed separately. 
With Eq.\eqref{imageJacobianMatrix_small}, the derivatives of $V_x$ and $V_y$ are
\begin{equation}\label{derived_Lyapunov}
	\begin{aligned}
		\frac{\partial V_x}{\partial{^i}\overline{p}_x}{^i}\dot{\overline{p}}_x &= {^i}\overline{p}_x\left(-\frac{{^g}v_x}{c_z} + {^g}v_z\frac{{^i}\overline{p}_x}{c_z} - (1+{^i}\overline{p}_x^2){^g}\omega_y\right),\\
		\frac{\partial V_y}{\partial{^i}\overline{p}_y}{^i}\dot{\overline{p}}_y &= {^i}\overline{p}_y\left(-\frac{{^g}v_y}{c_z} + {^g}v_z\frac{{^i}\overline{p}_y}{c_z} - {^i}\overline{p}_x{^i}\overline{p}_y{^g}\omega_y\right).\\
	\end{aligned}
\end{equation}
Let
\begin{equation}
	\begin{aligned}
		D_x({^i}\overline{p}_{x},{^g}v_x, {^g}v_z, c_z, {^g}\omega_y) &= \frac{\partial V_x}{\partial{^i}\overline{p}_x}{^i}\dot{\overline{p}}_x,\\
		D_y({^i}\overline{p}_{y},{^g}v_y, {^g}v_z, c_z, {^g}\omega_y) &= \frac{\partial V_y}{\partial{^i}\overline{p}_y}{^i}\dot{\overline{p}}_y.
	\end{aligned}
\end{equation}

\subsubsection{Searching datasets satisfies Lyapunov conditions}
%数据集需要满足的条件\eqref{searchingdatacondition}在这里实现为
The conditions that the datasets need to satisfy are shown in Eq. \eqref{searchingdatacondition}, where
\begin{equation}\label{Lyapunov_condition}
	\begin{aligned}
		W({^i}\overline{p}_x) &= \|{^i}\overline{p}_x\|^2,\\  W({^i}\overline{p}_y) &= \|{^i}\overline{p}_y\|^2,\\
	\end{aligned}
\end{equation}
and $\eta = 2$.
%\begin{equation}\label{Lyapunov_condition}
%	\begin{aligned}
%		D_x({^i}\overline{p}_{x},{^g}v_x, {^g}v_z, c_z, {^g}\omega_y) &\leqslant -\eta\|{^i}\overline{p}_x\|^2\\
%		D_y({^i}\overline{p}_{y},{^g}v_y, {^g}v_z, c_z, {^g}\omega_y) &\leqslant -\eta\|{^i}\overline{p}_y\|^2\\
%	\end{aligned}
%\end{equation}
The two groups of datasets are expressed as
\begin{equation}
	\begin{aligned}
		\mathcal{S}_x &= \left\{({^i}\overline{p}_{x,j}, {^g}v_{x,j}, {^g}v_{z,j}, c_{z,j}, {^g}\omega_{y,j}),j=1,\cdots,N\right\},\\ 
		\mathcal{S}_y &= \left\{({^i}\overline{p}_{y,j}, {^g}v_{y,j}, {^g}v_{z,j}, c_{z,j}, {^g}\omega_{y,j}),j=1,\cdots,N\right\}.\\
	\end{aligned}
\end{equation}
The RoI of the state in this study are shown in Tab.\ref{ROIrange}, where ${^g}w_y$ is considered to be small value. The interception task is mainly accomplished through the translational movement of the multicopter.
\begin{table}[!t]
	\centering
	\caption{The RoI of system state and system input}
	\begin{tabular}{c|c|c}
	\hline
	State & Range & Unit\\
	\hline
	${^i}\overline{p}_x$ & $\left[-1\ 1\right]$ & - \\
	${^i}\overline{p}_y$ & $\left[-1\ 1\right]$ & -\\
	${^g}\omega_y$ & $\left[-0.2\ 0.2\right]$ & rad/s\\
	${^g}v_z$ & $\left[0.1\ 15\right]$ & m/s\\
	$c_z$ & $\left[0.5\ 50\right]$ & m\\
	\hline
	\end{tabular}
	\label{ROIrange}
\end{table}
%可采用均匀遍历的方式或者随机数的方式查找符合条件\eqref{Lyapunov_condition}的数据点来构造数据集$\mathcal{S}_x$和$\mathcal{S}_y$

\subsubsection{Training neural network $c_{\theta_{x,k}}$ and $c_{\theta_{y,k}}$}
The neural network policies $c_{\theta_{x,k}}$ and $c_{\theta_{y,k}}$ are trained by using the datasets $\mathcal{S}_x$ and $\mathcal{S}_y$, and by solving the optimization problems
\begin{equation}\label{castfcn_controller_in_practice}
	\begin{aligned}
		L(\theta_{x,k}) &= \min_{c_{{\theta}_{x,k}}} \frac{1}{N}\sum_{j=1}^{N}\|c_{{\theta}_{x,k}}\left({^i}\overline{p}_{x,j}, {^g}v_z, c_z\right)-{^g}v_{x,j}\|,\\
		L(\theta_{y,k}) &= \min_{c_{{\theta}_{y,k}}} \frac{1}{N}\sum_{j=1}^{N}\|c_{{\theta}_{y,k}}\left({^i}\overline{p}_{y,j}, {^g}v_z, c_z\right)-{^g}v_{y,j}\|.\\
	\end{aligned}
\end{equation}
%\begin{equation}
%	\begin{aligned}
%		\min_{c_{x,\mathrm{net}},\eta\in\mathbb{R}} & -\eta\\
%		\mathrm{s.t.} & -D_x({^i}\overline{p}_{x,j},c_{t,x,\mathrm{net}}({^i}\overline{p}_{x,j})) -\eta\|{^i}\overline{p}_{x,j}\|^2 \geqslant 0
%	\end{aligned}
%\end{equation}
%\begin{equation}\label{castfcn_controller_in_practice}
%	\begin{aligned}
%		L_c &= \lambda_1 \max(D(x_j,c_{\theta_k}({^i}\overline{p}_{x,j})),0)\\
%		&+ \frac{\lambda_2}{N}\sum_{j=1}^{N}(D(x_j,c_{\theta_k}({^i}\overline{p}_{x,j}))) + \lambda_3\|\theta_k - \theta_{k-1}\|^2\\
%	\end{aligned}
%\end{equation}

%文献\cite{NeuralLyapunovControl2019}提到对于控制问题不应采用ReLu激活函数，而是应该使用可微的Tanh函数作为激活函数，防止训练过程中出现突变的梯度。
% TODO 本方法与D学习方法的区别对比表格
% TODO 理论证明

\subsection{Stability verification}
%TODO 控制结构图
This subsection encompasses the validation of the almost Lyapunov condition for the trained neural network policy and the outcomes of the simulation utilizing the simplified model.
\begin{figure}[t]
	\centering
	%  \framebox{\parbox{3.3in}{\includegraphics[width=8.3cm]{img/ContourPlot.pdf}}}
	\includegraphics[width=8.5cm]{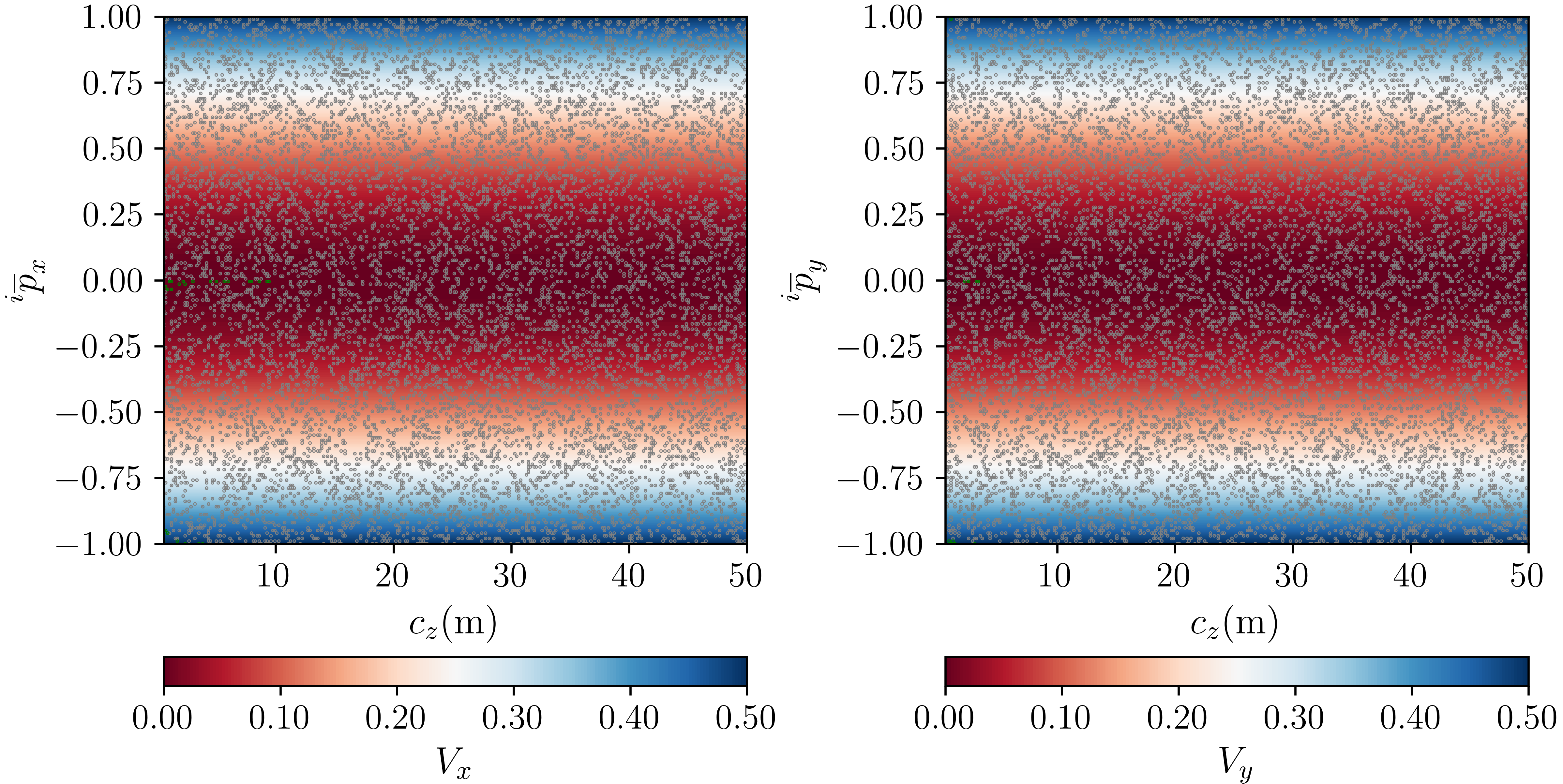}
	\caption{The region of attraction for the interception system with the trained neural network policy when ${^g}v_z = 15\mathrm{m/s}$ and ${^g}\omega_y = 0\mathrm{rad/s}$. Grey points denote negative values, while green points signify positive values.} 
	\label{train_result_simple_sim_point}
\end{figure}
\subsubsection{Verifying almost Lyapunov conditions}
Before deploying the trained neural network policy, its stability is verified using the almost Lyapunov condition\cite{LiuAlmostLyapunovfunctions2020}, which is used for stability analysis in \cite{ChangStabilizingNeuralControl2021,DawsonSafeNonlinearControl2022}.
%This theorem proves stability by allowing a small and finite set of violation states where the derivative of Lyapunov is positive.
The verification of the almost Lyapunov conditions is achieved by the sample-based method.
%需要我们首先在ROI区间对状态进行重新密集采样构造状态数据集，在数据集基础上，计算Lyapunov函数的导数
First, we need to intensively resample the states within the RoI as shown in Tab.\ref{ROIrange} to construct state datasets. Based on this datasets, we then calculate the derivative of the Lyapunov function:
\begin{equation}\label{Dfunctioninsgn}
	\begin{aligned}
		\mathrm{sgn}\left(D_x\left({^i}\overline{p}_x,c_{\theta_{x,k}},{^g}v_z,c_z,{^g}\omega_y\right)\right),\\
		\mathrm{sgn}\left(D_y\left({^i}\overline{p}_y,c_{\theta_{y,k}},{^g}v_z,c_z,{^g}\omega_y\right)\right),\\
	\end{aligned}
\end{equation}
where $\mathrm{sgn}$ is the sign function. A negative value of this function indicates that the closed-loop system under this state tends towards the desired state.

%式\eqref{Dfunctioninsgn}的结果取决于多个状态输入。为了能够便于展示验证结果，这里固定一些相对不重要的状态，即
The results of Eq.\eqref{Dfunctioninsgn} are decided by multiple state inputs. To facilitate the presentation and verification of the results, some relatively unimportant states are fixed here.
Let ${^g}v_z = 15\mathrm{m/s}$ and ${^g}\omega_y = 0\mathrm{rad/s}$. 
%\begin{table}[!t]
%	\centering
%	\caption[Init state table]{The init states in Fig.\ref{train_result_simple_sim}}
%	\label{table_init_state}
%	\begin{tabular}{|c|c|c|c|c|c|c|}
%		\hline
%		&\begin{tabular}[c]{@{}c@{}}Init\\ state 1\end{tabular}&\begin{tabular}[c]{@{}c@{}}Init\\ state 2\end{tabular}&\begin{tabular}[c]{@{}c@{}}Init\\ state 3\end{tabular}&\begin{tabular}[c]{@{}c@{}}Init\\ state 4\end{tabular}&\begin{tabular}[c]{@{}c@{}}Init\\ state 5\end{tabular}&\begin{tabular}[c]{@{}c@{}}Init\\ state 6\end{tabular}\\
%		\hline
%		$c_z$&30&40&50&25&35&45\\
%		${^i}\overline{p}_x$&0.05&0.5&0.85&-0.35&-0.7&-0.9\\
%		${^i}\overline{p}_y$&-0.5&-0.75&-0.5&0.2&0.6&0.8\\
%		\hline
%	\end{tabular}
%\end{table}
Fig.\ref{train_result_simple_sim_point} presents the domain of attraction of the interception system when using the trained policies $c_{\theta_{x,k}}$ and $c_{\theta_{y,k}}$, where $k=5$. The values of $V_x$ and $V_y$ are distinguished by using different shades of background color. The positive values of $D_x$ and $D_y$ are denoted by green points, while the negative values are denoted by grey points.
%从左图中可以得知，只有在$c_z<10\mathrm{m}$且${^i}\overline{p}_x=0$附近时，存在$D_x>0$的情况。这表明在大部分情况下$V_x$趋向于0，也就是${^i}\overline{p}_x$趋向于0。对于右图也是同理。
From the left figure, it can be seen that only when $c_z < 10\mathrm{m}$ and ${^i}\overline{p}_x = 0$, there is a situation where $D_x > 0$. This indicates that in most cases $V_x$ tends to 0, that is, ${^i}\overline{p}_x$ tends to 0. The same is true for the right figure.
%从拦截效果的角度分析，这意味着飞行器在与目标逼近的时候，目标中心点与图像中心点会存在静态误差。这是因为，在${^i}\overline{p}_x=0$附近存在$D_x$大于0的情况，而${^i}\overline{p}_x$会稳定收敛到$D_x = 0$的位置，此时${^i}\overline{p}_x\neq0$。
This means that when the aircraft approaches the target, there will be a static error on ${^i}\overline{p}_x$ and ${^i}\overline{p}_y$. This is because, near ${^i}\overline{p}_x = 0$, there are cases where $D_x>0$ and $D_y>0$. Then, ${^i}\overline{p}_x$ and ${^i}\overline{p}_y$ will converge to the position where $D_x = 0({^i}\overline{p}_x \neq0)$ and $D_y = 0({^i}\overline{p}_y \neq0)$, respectively.

%TODO 改成更加通用的像素距离。
\begin{remark}
	%由于目标并非质点，即便${^i}\overline{p}_x\neq0$，只要${^i}\overline{p}_x$收敛到足够接近0的位置依旧实现目标拦截。显然，能够拦截目标所允许的$\left|{^i}\overline{p}_x\right|$最大值与目标自身的尺寸有关系。
	Given that the target is not a point mass, interception can still be accomplished even if ${^i}\overline{p}_x \neq 0$, provided that ${^i}\overline{p}_x$ converges to a location sufficiently near 0. The greatest permissible value of $\left|{^i}\overline{p}_x\right|$ for successfully intercepting the target is contingent upon the dimensions of the target itself.
	%考虑更一般的情况，定义目标在图像坐标系中距离坐标系中心的像素距离为${^i}{p} = \sqrt{{^i}{p}_x^2 + {^i}{p}_y^2}$，
	
	Considering a more general case, the pixel distance of the target from the center of the image coordinate system is defined as 
	\begin{equation}
		{^i}{p} = \sqrt{{^i}{p}_x^2 + {^i}{p}_y^2}.
	\end{equation}
	%如图\ref{maxpx}所示，假设物距远大于焦距，视距近似等于焦距，则存在以下关系
	As shown in Fig.\ref{maxpx}, assuming that the object distance is much greater than the focal length, the image distance is approximately equal to the focal length. Then the following relationship exists:
	\begin{equation}
		\frac{r}{c_z} = \frac{{^i}p_{\max}}{f_{\mathrm{oc}}} = {^i}\overline{p}_{\max}
	\end{equation}
	%其中，$r$表示被拦截目标轮廓内切圆的半径，${^i}p_{\max}$表示能够拦截目标所允许的${^i}p$的最大值（假设目标静止），${^i}\overline{p}_{\max}$为${^i}p_{\max}$的归一化值。令$c_z = 1\mathrm{m}$且满足关系$c_z\gg f_\mathrm{oc}$，就有
%	Here, $r$ represents the radius of the inscribed circle of the intercepted target's contour, ${^i}p_{\max}$ represents the maximum allowable ${^i}p$ for intercepting the target (assuming the target is stationary), and ${^i}\overline{p}_{\max}$ is the normalized value of ${^i}p_{\max}$. 
	In this context, $r$ denotes the radius of the inscribed circle of the intercepted target's contour, ${^i}p_{\max}$ signifies the maximum permissible ${^i}p$ for target interception (assume the target remains stationary), and ${^i}\overline{p}_{\max}$ represents the normalized value of ${^i}p_{\max}$.

	Let $c_z = 1\mathrm{m}$ and satisfy the relationship $c_z \gg f_\mathrm{oc}$, then it has
	\begin{equation}\label{key}
		{^i}\overline{p}_{\max} = r
	\end{equation}
	
	%式\eqref{key}表明，当相机与目标物距为1m时，如果目标在相机图像坐标系中距离坐标系中心的距离${^i}\overline{p}\leqslant r$，那么可以认为此时飞行器继续沿着光轴当前指向继续移动能够命中目标。
	Eq.\eqref{key} denotes that when the distance from the camera to the target object is one meter, if the target's distance from the center of the camera's image coordinate system, ${^i}\overline{p}$, is less than or equal to $r$, it can be inferred that if the aircraft persists in its current trajectory along the optical axis, it will successfully touch the target.
	\begin{figure}[!t]
		\centering
		\includegraphics[width=8.5cm]{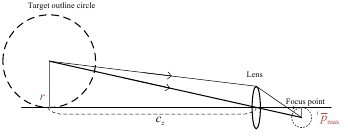}
		\caption{Assuming the target is stationary, to achieve target interception, the relationship between the target size and the distance of the target from the center of the camera image.}
		\label{maxpx}
	\end{figure}
\end{remark}
%事实上，随着飞行器持续接近目标，单位像素偏移对应的实际偏移距离越小

%（但是实际上，$\overline{p}_x$一直没有收敛到非常接近0的位置，但是就实际情况来看，即便$\overline{p}_x$一直保持较小的值，如0.1也能集中目标，因为在非常接近目标的时候即使微小的偏移也会导致$\overline{p}_x$变得很大，这点建议通过理论公式来说明，证明对于多大的目标，在距离目标1m的时候若能够击中目标，目标体积和${^i}\overline{p}_x$容忍范围的数学关系）

%我们采用简化模型进行仿真以直观地对控制器进行验证。
\subsubsection{Simplified simulation}
To verify the policies, we employed a simplified simulation model assuming instantaneous execution of velocity commands by the multicopter. The yaw rate ${^g}\omega_y$ was set to 0. Based on Eq.\eqref{small_interception_model}, the simulation model was simplified to Eq.\eqref{partcle_sim_model}:
%We used a simplified model for simulation to verify the controller.
%Assume the velocity command is executed instantaneously by the multicopter.
%且偏航角速率${^g}\omega_y$设置为0. 根据式\eqref{small_interception_model}，仿真模型改写为\eqref{partcle_sim_model}
\begin{equation}
	\label{partcle_sim_model}
		\begin{bmatrix}
			\dot{c}_z\\
			{^i}\dot{\overline{p}}_x\\
			{^i}\dot{\overline{p}}_y\\
		\end{bmatrix} = \begin{bmatrix}
		0 &0 &-1\\
		-\frac{1}{c_z} & 0 & \frac{{^i}\overline{p}_x}{c_z}\\
		0 & -\frac{1}{c_z} & \frac{{^i}\overline{p}_y}{c_z}\\
		\end{bmatrix}\begin{bmatrix}
		{^g}v_x\\
		{^g}v_y\\
		{^g}v_z\\
		\end{bmatrix},
\end{equation}
where ${^g}v_x$ is determined by the neural network $c_{\theta_{x,k}}$, while ${^g}v_y$ is generated by $c_{\theta_{y,k}}$. The value of ${^g}v_z$ is fixed at 15m/s. Simulations were conducted using six distinct sets of initial states. 
%暂时取消该表格，用处不大as specified in Tab.\ref{table_init_state}.
%其中，${^g}v_x$由神经网络$c_{\theta_{x,k}}$给出，${^g}v_y$由神经网络$c_{\theta_{y,k}}$给出。同时，令沿视轴速度${^g}v_z = 15\mathrm{m}/\mathrm{s}$。
%使用六组不同的初始状态进行仿真，初始状态见表\ref{table_init_state}.

\begin{figure}[!t]
	\centering
	\includegraphics[width=8.5cm]{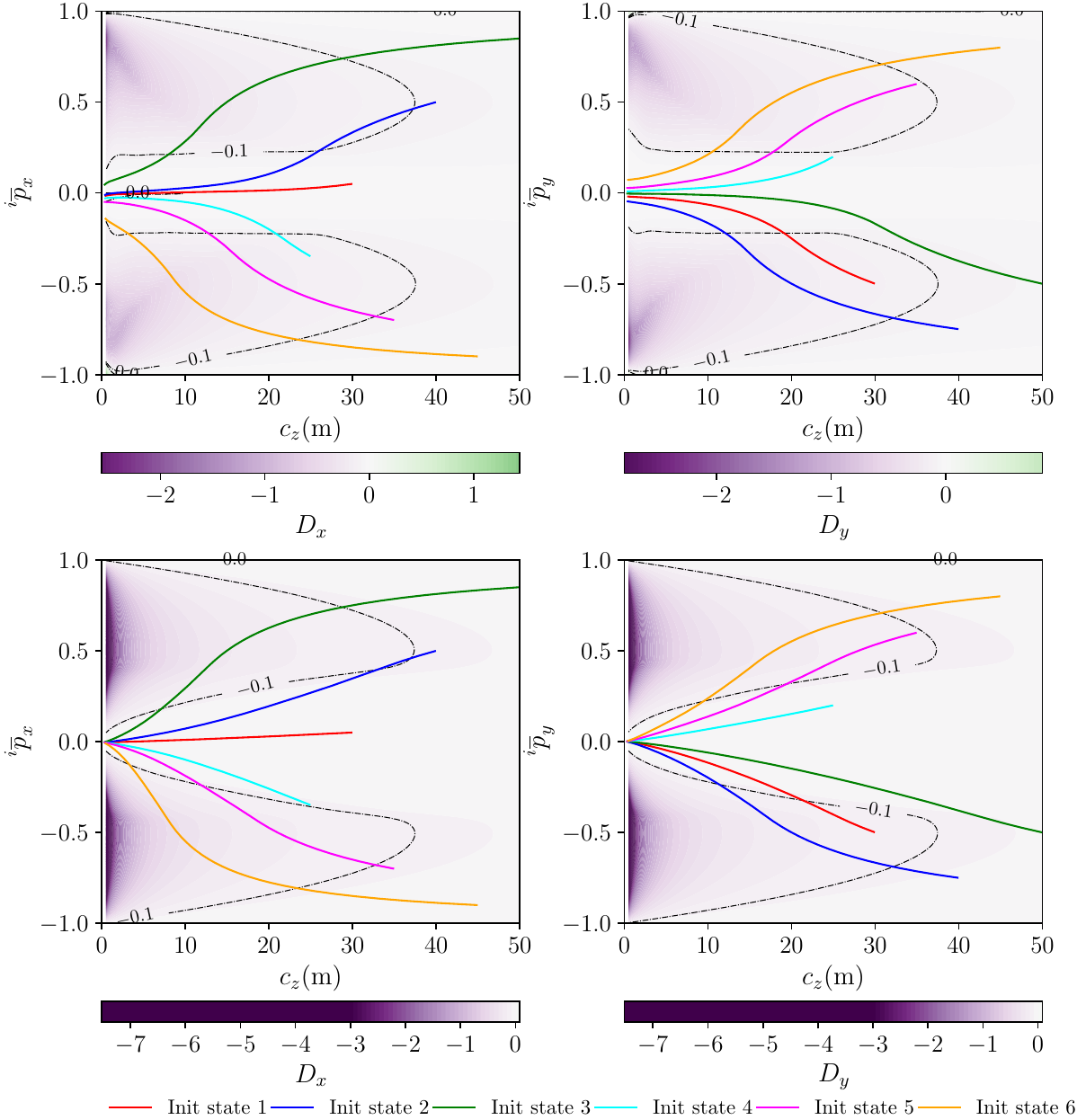}
	\caption{The two top plots show the flight trajectories in the simulation using model \eqref{partcle_sim_model} with ${^g}v_z = 15\mathrm{m/s}$ and ${^g}\omega_y = 0\mathrm{rad/s}$, the two bottom plots show the flight trajectories in the same simulation but with $c_z \equiv 10\mathrm{m}$ for the policies $c_{\theta_{x,k}}$ and $c_{\theta_{y,k}}$.}
	\label{train_result_simple_sim}
\end{figure}

The two top plots in Fig.\ref{train_result_simple_sim} present the simulation results under different initial states. Purple regions indicate coordinates where $D_x<0$ and $D_y<0$, with darker shades corresponding to smaller (more negative) $D$ values. Conversely, green denotes $D > 0$. States ${^i}\overline{p}_x$ and ${^i}\overline{p}_y$ exhibit faster decay rates (indicating accelerated convergence) in darker purple regions.
Furthermore, as $c_z \rightarrow 0$, the convergence ${^i}\overline{p}_x \rightarrow 0$ and ${^i}\overline{p}_y \rightarrow 0$ hold.

%图\ref{train_result_simple_sim}展示了不同初始状态下的仿真结果。其中，紫色用于表示所对应的坐标点状态的$D$值为负，颜色越深则$D$值越小。相反的是，绿色表示$D$值为正。可以观察到，状态${^i}p_x$和${^i}p_y$在紫色背景加深的地方衰减速度更快，即收敛速度越快。
%同时，随着$c_z\rightarrow0$，${^i}\overline{p}_x\rightarrow0$和${^i}\overline{p}_y\rightarrow0$是成立的。

In the design of subsection \textit{Training of interception policy}\ref{training_of_controller}, the neural network policies $c_{\theta_{x,k}}$ and $c_{\theta_{y,k}}$ require the value of the object distance $c_z$ to obtain the speed commands ${^g}v_x$ and ${^g}v_y$.
In practice, obtaining $c_z$ requires the design of additional algorithms. 
Here, we attempt to set a fabricated object distance, $c_z = 10\mathrm{m}$, for the policies for further verification.
As shown in the two bottom plots in Fig.\ref{train_result_simple_sim}, as $c_z \rightarrow 0$, ${^i}\overline{p}_x \rightarrow 0$ and ${^i}\overline{p}_y \rightarrow 0$ still hold. 
The two bottom plots in Fig.\ref{train_result_simple_sim} provide better convergence results.
By comparing the distribution of $D$ values and the trend of state trajectories in these two figures and combining with the constraint conditions \eqref{searchingdatacondition}, it can be seen that setting a larger $\eta$ or designing more reasonable constraint conditions is useful for improving the convergence speed and steady-state error of the state. 
It should be noted that in the simulation, the dynamic process required for the quadcopter to execute the speed command is not considered. 
Therefore, in the training process, $\eta$ cannot be unlimited and set to larger values. 
A more reasonable approach is to improve the constraint conditions \eqref{searchingdatacondition} of the datasets.

%在Section \ref{training_of_controller}的设计中，神经网络控制器$c_{\theta_{x,k}}$和$c_{\theta_{y,k}}$需要物距$c_z$的值以得到速度指令${^g}v_x$和${^g}v_y$.
%在实际中获取$c_z$需要设计额外的算法。这里我们尝试给控制器设置伪造的物距$c_z = 10\mathrm{m}$，来进一步验证控制器的仿真效果。
%如图\ref{train_result_simple_sim_fixdis}所示，随着$c_z\rightarrow0$, ${^i}\overline{p}_x\rightarrow0$和${^i}\overline{p}_y\rightarrow0$依旧成立。相比图\ref{train_result_simple_sim}，图\ref{train_result_simple_sim_fixdis}给出了更好的收敛结果。
%对比这两幅图中的$D$值分布以及状态轨迹的变化趋势，并结合本研究采用的约束条件\eqref{searchingdatacondition}，可以看出在训练过程设置更大的$\eta$或者设计更合理的约束条件，可用于改进状态的收敛速度和稳态误差。需要注意的是，在本节所提到的仿真，没有考虑飞行器执行速度指令所需要的动态过程，所以在训练过程中$\eta$并不能无限制的取更大的数值，更合理的方式是改进数据集约束条件\eqref{searchingdatacondition}。
%\begin{figure}[!t]
%	\centering
%	\includegraphics[width=8.5cm]{img/SimpleSimFlightTrajectoryFixDistance.pdf}
%	\caption{Flight trajectory simulation using \eqref{partcle_sim_model} and set $c_z \equiv 10\mathrm{m}$ for the policies $c_{\theta_{x,k}}$ and $c_{\theta_{y,k}}$.}
%	\label{train_result_simple_sim_fixdis}
%\end{figure}

%\begin{figure}[!t]
%	\centering
%	%  \framebox{\parbox{3.3in}{\includegraphics[width=8.3cm]{img/ContourPlot.pdf}}}
%	\includegraphics[width=8.0cm]{img/ContourPlotFixDistance.pdf}
%	\caption{Caption}
%	\label{train_result_simple_sim_fixdistance}
%\end{figure}

% TODO:增加代价函数减小的过程。

\section{EXPERIMENT}
%本节将给出所训练算法的部署运行效果，以说明本文提出的算法是有效的。首先，介绍了用于验证算法的飞行平台构成，和控制结构。
%随后，给出基于RflySim\cite{DaiRFlySimAutomatictest2021,DaiSimulationCredibilityAssessment2021,WangRflySimRapidMulticopter2021}平台的硬件在环仿真结果。最后展示真实世界下的试验结果。
%
%
\subsection{Quadcopter setup}
%我们采用四旋翼飞行器作为算法验证平台。神经网络控制器和用于目标识别的算法运行在机载计算机上，机载计算机采用NVIDIA Jetson Xavier NX 板卡。神经网络控制器以30Hz运行。采用Yolo-v5进行目标识别，并使用红色气球模拟被拦截的目标。相机图像输出规格为1280x720@10Hz，目标识别算法以10Hz运行。飞行器的姿态控制任务和响应机载计算机速度指令的任务由开源飞控PX4/Pixhawk完成。图\ref{quadcopter}给出了飞行器实物图。

%采用四旋翼飞行器平台验证算法，搭载NVIDIA Jetson Xavier NX机载计算机。神经网络控制器（30Hz）与目标识别算法（Yolo-v5, 10Hz）部署于该平台，以红色气球模拟拦截目标。相机输入为1280x720@10Hz。由PX4/Pixhawk飞控负责姿态控制并对速度指令进行响应。飞行器实体见图\ref{quadcopter}。

We validated algorithms on a quadcopter featuring NVIDIA Jetson Xavier NX, running the neural network policies (30 Hz) and a YOLOv5 target recognition instance (on red balloons) with 1280×720@10 Hz camera input. The attitude control was handled by the PX4/Pixhawk system for flight control, which responded to velocity commands.  

As shown in Fig.\ref{quadcopter}, the quadcopter has a total mass of $1.1\mathrm{kg}$. Each rotor has a time constant of $0.03\mathrm{s}$ and can generate a maximum thrust of $11.5\mathrm{N}$. The distance from the center of the vehicle to each rotor, known as the arm length, is $0.20\mathrm{m}$. Its moments of inertia are $0.0076\mathrm{kg}\cdot \mathrm{m}^2$ about the ${^b}x$-axis, $0.0089 \mathrm{kg}\cdot \mathrm{m}^2$ about the ${^b}y$-axis, and $0.0137 \mathrm{kg}\cdot \mathrm{m}^2$ about the ${^b}z$-axis.

Unlike the simulation based on Eq.\eqref{partcle_sim_model}, the gimbal camera's yaw rate ${^g}\omega_y$ in physical experiments was directly controlled by proportional feedback of ${^i}\overline{p}_x$. Consequently, the four control inputs driving the quadcopter are
\begin{equation}
	\begin{aligned}
		{^g}v_x &= c_{\theta_{x,k}},\\ 
		{^g}v_z &= 15,\\
		{^g}v_y &= c_{\theta_{y,k}},\\ 
		{^g}\omega_y &= 0.002\cdot{^i}\overline{p}_x.\\
	\end{aligned}
\end{equation}
%对于PX4飞控而言，需要基于飞行器当前的偏航角，将${^g}v_x$、${^g}v_y$、${^g}v_z$转换到惯性坐标系下。
The velocity commands ${^g}v_x$, ${^g}v_y$, and ${^g}v_z$ must be transformed into EFCS based on the current yaw angle before sending to the PX4 flight controller.

%图\ref{interceptionprocess}给出了飞行过程中捷联摄像头获取的视频截图，涵盖了四旋翼刚进入拦截飞行状态到即将命中目标时的瞬间图像。可以看到在刚进入拦截状态时由于俯仰姿态变化，目标出现在图像上侧边缘。在随后的图像中，目标逐步趋于图像中心。
%图\ref{realflightcmdrespond}给出了机载计算机向飞控发出的控制指令，以及飞行器对控制指令的响应结果。
%图\ref{TargetareaandImageerror}给出了在拦截过程中，目标在图像中的矩形框面积（对应图\ref{interceptionprocess}中红框大小）的变化趋势和目标中心在图像坐标系中坐标的变化。 
%图\ref{position}则给出了拦截过程中，飞行器位置的散点图。
\subsection{Experiment result}
\begin{figure}[!t]
	\centering
	\includegraphics[width=8.5cm]{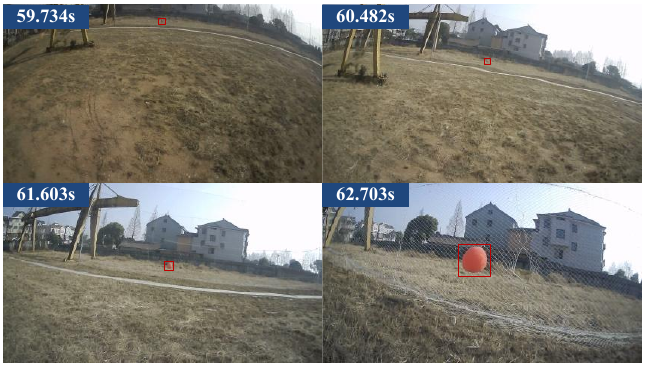}
	\caption{Video screenshots from the strapdown camera of the quadrotor during the experiment. Each screenshot is labeled with the video timestamp.}
	\label{interceptionprocess}
\end{figure}

\begin{figure}[!t]
	\centering
	\includegraphics[width=8cm]{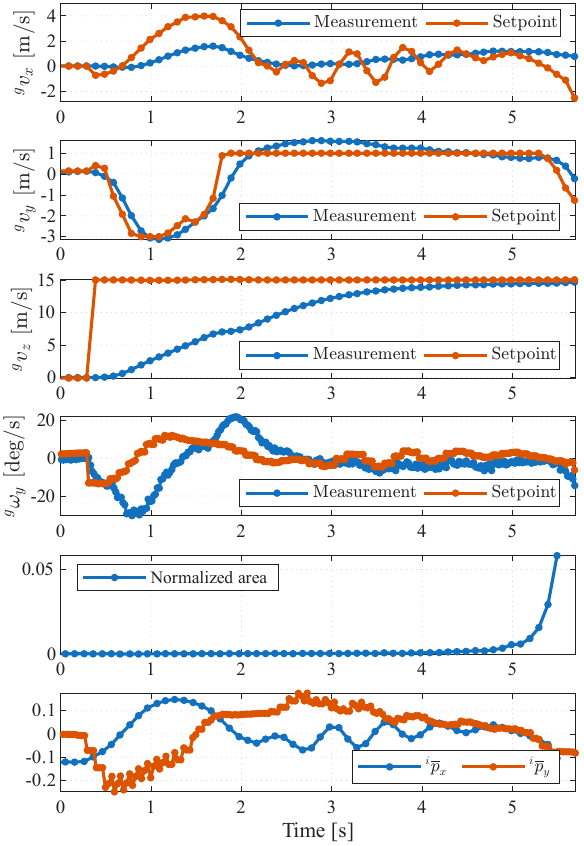}
	\caption{The response of the quadrotor to the command from the onboard computer}
	\label{realflightcmdrespond}
\end{figure}

\begin{figure}[!t]
	\centering
	\includegraphics[width=8cm]{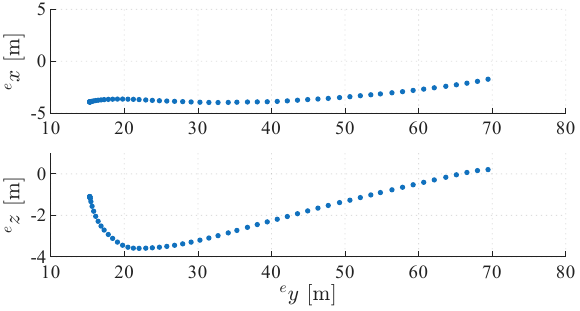}
	\caption{Scatter trajectory of the quadrotor during interception}
	\label{position}
\end{figure}

Fig.\ref{interceptionprocess} presents the video snapshots captured by the strapdown camera during the flight, covering the images from the moment the quadcopter just entered the interception flight state until the instant it was about to hit the target. When the interception state was just entered, due to the change in pitch attitude, the target appeared on the upper edge of the image. In the subsequent images, the target gradually moved towards the center of the image.
%Fig. \ref{interceptionprocess} presents video stills captured by the strapdown camera during flight, showing the interception sequence from initial engagement to imminent target impact. In the initial interception phase (Fig. \ref{interceptionprocess}a), the target appears near the upper image edge due to pitch attitude changes. Subsequent frames show the target progressively converging toward the image center. 
The first four subplots in Fig.\ref{realflightcmdrespond} show velocity commands issued by the onboard computer and the actual response. The last two subplots in Fig.\ref{realflightcmdrespond} track two metrics during interception: i) the bounding box area of the target in image space (represented by red boxes in Fig.\ref{interceptionprocess}), and ii) the values of ${^i}\overline{p}_x$ and ${^i}\overline{p}_y$. Fig.\ref{position} shows the scatter plot of the quadcopter's position during the interception.
%可以看出，飞行器沿着${^g}x$-axis（水平方向）的速度指令存在一定波动，即在飞行过程中飞行器存在一定左右摇摆的情况，这与图\ref{realflightcmdrespond}中${^i}\overline{p}_x$曲线的变化对应。
%而飞行器沿着${^g}y$-axis（垂直方向）的速度指令在拦截任务起初有3m/s的波动，这是飞行器在通过俯仰姿态变化实现前飞时目标在图像中位置突变导致的。根据飞行器姿态角和相机安装角将目标在捷联相机图像中的位置等效转换为其在云台相机图像中后，其位置与飞行器姿态变化实现了解耦。然而，实际中的图像采集和姿态角解算没有严格同步开展，这就导致转换的结果依旧会受到飞行器姿态角的轻微影响，这与图\ref{realflightcmdrespond}中${^i}\overline{p}_y$曲线的变化对应。
%飞行器沿着${^g}z$-axis的速度${^g}v_z$逐步加速至$15$m/s，这对应了图\ref{position}中散点间距逐步增大。
The ${^g}x$-axis velocity command exhibits fluctuations, causing lateral sway during flight, correlating with ${^i}\overline{p}_x$ variations in Fig.\ref{realflightcmdrespond}.
Along the ${^g}y$-axis, an initial 3 m/s fluctuation occurs due to abrupt target displacement when pitching forward. Although strapdown-to-gimbal coordinate conversion decouples the target position from the attitude, residual attitude sensitivity persists from unsynchronized image capture and attitude computation. This aligns with ${^i}\overline{p}_y$ dynamics in Fig.\ref{realflightcmdrespond}.
%${^g}v_z$ ramps to 15m/s, reflected by increasing scatter point spacing in Fig. \ref{position}.

%TODO 下面这段为了节省篇幅省略了
%在前面的稳定性验证中，我们没有考虑偏航角速度的变化，从而更多地关注${^i}\overline{p}_x$和${^i}\overline{p}_y$的变化对系统稳定性的关系。从试验结果中可以看出，飞行器的偏航角速度在拦截过程起初2.5s内出现了较大的控制量, 随后稳定在$\pm 5$deg/s附近。这说明稳定性分析中对偏航角速率影响的忽略并没有对实际部署结果产生致命的影响。
In the previous stability verification, we did not take into account the changes in yaw angular velocity, thus focusing more on the relationship between the changes of ${^i}\overline{p}_x$(${^i}\overline{p}_y$) and the system stability. 
%According to the experiment, it can be seen that the yaw angular velocity of the quadcopter reached a large control amount in the first 2.5s of the interception process, and then stabilized around $\pm 5$ deg/s. 
The experiment indicates that the quadcopter's yaw angular velocity attained a significant control magnitude during the initial 2.5 seconds of the interception process, subsequently stabilizing at $\pm 5$ deg/s.
This indicates that the neglect of the influence of yaw angular rate in the stability verification did not have a fatal impact on the actual deployment results.
%TODO 上面这段为了节省篇幅省略了

%\subsection{Hardware in the loop simulation}
%我们采用RflySim\cite{DaiRFlySimAutomatictest2021,DaiSimulationCredibilityAssessment2021,WangRflySimRapidMulticopter2021}开展硬件在环仿真验证。具体来说，四旋翼仿真模型和虚拟场景由个人计算机完成，飞控通过串口与计算机中的四旋翼飞行器模型进行通信，同时虚拟场景中可以为四旋翼飞行器挂在虚拟摄像头，并通过以太网络将摄像头采集的图像信息发送给机载计算机，机载计算机运行目标识别算法和神经网络控制器后，将速度指令通过串口发送给飞控。
%
%\begin{itemize}
%	\item 速度指令vs速度
%	\item ${^i}\overline{p}_x$和${^i}\overline{p}_y$
%	\item 飞行轨迹，水平和高度
%	\item 画面九宫格
%\end{itemize}
%\subsection{Real flight experiment}

%\begin{figure}[!t]
%	\centering
%	\includegraphics[width=8cm]{img/TargetareaandImageerror.pdf}
%	\caption{The bounding box area and the normalized position of the target}
%	\label{TargetareaandImageerror}
%\end{figure}

%总之，基于所提方法训练的拦截控制器在实际部署中能够最终以设计速度完成拦截飞行任务。这进一步表明，基于本方法，能够在不依赖真实数据的情况下，仅基于先验知识所建立的系统模型人工构造数据集，即可通过离线训练快速得到初始可用的拦截控制器。本文所提的方法有望拓展到其他应用场景。
The trained interception control policies successfully executed missions at target speeds during deployment. This confirms our method's ability to rapidly develop the mediocre control policy via offline training using only posteriori knowledge and synthetic datasets without real flight data. 
%TODO 是否要提及：The approach shows potential for broader applications.
%结果表明，基于先验知识所建立的系统模型生成符合稳定性约束数据集所训练得到控制器可基本完成拦截任务。

%\begin{table}[h]
%\caption{An Example of a Table}
%\label{table_example}
%\begin{center}
%\begin{tabular}{|c||c|}
%\hline
%One & Two\\
%\hline
%Three & Four\\
%\hline
%\end{tabular}
%\end{center}
%\end{table}

\section{CONCLUSIONS AND FUTURE WORK}

%本文就多旋翼飞行器拦截问题，研究了一种不依赖真实飞行数据，而采用数学模型构造数据集实现控制器快速训练的方法。相比于现有的Lyapunov学习方法聚焦于对初始可用控制器的迭代优化，本方法聚焦于初始控制器的快速训练生成，用于替代传统控制设计方法。
%仿真与试验表明了本方法的有效性。同时可以看到，本文在处理拦截问题时进行了简化，例如训练神经网络时忽略了云台相机的偏航运动，且在构造数据集的时候未考虑飞行器响应速度指令的动态过程。这些简化可能对真机试验效果产生负面影响，有必要在后续研究中，引入更全面的模型。本文仅基于前馈神经网络进行练，未来会结合更复杂的神经网络，如循环神经网络，对所提方法进行验证。
%This paper presents a data-free interception method for multicopters, using mathematical models to generate synthetic datasets for rapid controller training. Unlike existing Lyapunov learning methods that optimize initial controllers iteratively, our approach rapidly generates initial controllers, bypassing traditional control design. Simulations and experiments validate the method's effectiveness, though simplifications like ignoring gimbal yaw motion and actuator dynamics may limit real-world performance. Future work will incorporate comprehensive models and advanced neural networks (e.g., RNNs) to enhance robustness.

This paper presents a real data-free neural network policy training method, using mathematical models to generate synthetic datasets for the rapid policy training. 
Our method rapidly trains initial stable controllers, bypassing traditional control design, after which the previously learned controller may be repeatedly improved using RL and LLC methods.
Simulations and experiments of image-based visual servoing for the multicopter interception validate the method's effectiveness.
%, though simplifications like ignoring gimbal yaw motion and actuator dynamics may limit real-world performance.
Future work will incorporate comprehensive models and advanced neural networks (e.g., RNNs) to enhance robustness.

\bibliographystyle{IEEEtran}
\bibliography{root}

\end{document}